\documentclass[conference]{IEEEtran}
\usepackage{amsmath}
\usepackage{amssymb}
\usepackage{amsfonts}
\usepackage{graphicx}
\usepackage{epsfig}
\usepackage{subfigure}
\usepackage{psfrag}
\usepackage{cite}
\usepackage{latexsym}
\usepackage{url}
\usepackage{color}
\PassOptionsToPackage{bookmarks={false}}{hyperref}
\IEEEoverridecommandlockouts
\begin{document}
\title{Max-Min Fairness in IRS-Aided Multi-Cell MISO Systems via Joint Transmit and Reflective Beamforming
\thanks{This work was supported in part by the National Key R\&D Program of China (No. 2018YFB1800800), the Natural Science Foundation of China (Nos. 61871137 and 11671419), the Guangdong Province Key Area R\&D Program (No. 2018B030338001), and the Guangdong Province Basic Research Program (Natural Science) (No. 2018KZDXM028). J. Xu is the corresponding author.}}

\author{Hailiang Xie$^1$, Jie Xu$^1$, and Ya-Feng Liu$^2$\\
$^1$School of Information Engineering, Guangdong University of Technology, Guangzhou, China \\
$^2$LSEC, ICMSEC, AMSS, Chinese Academy of Sciences, Beijing, China\\
E-mail: hailiang.gdut@gmail.com,~jiexu@gdut.edu.cn,~yafliu@lsec.cc.ac.cn
}
\setlength\abovedisplayskip{2.5pt}
\setlength\belowdisplayskip{2.5pt}
\maketitle

\begin{abstract}
This paper investigates an intelligent reflecting surface (IRS)-aided multi-cell multiple-input single-output (MISO) system consisting of several multi-antenna base stations (BSs) each communicating with a single-antenna user, in which an IRS is dedicatedly deployed for assisting the wireless transmission and suppressing the inter-cell interference. Under this setup, we jointly optimize the coordinated transmit beamforming at the BSs and the reflective beamforming at the IRS, for the purpose of maximizing the minimum weighted received signal-to-interference-plus-noise ratio (SINR) at users, subject to the individual maximum transmit power constraints at the BSs and the reflection constraints at the IRS. To solve the difficult non-convex minimum SINR maximization problem, we propose efficient algorithms based on alternating optimization, in which the transmit and reflective beamforming vectors are optimized in an alternating manner. In particular, we use the second-order-cone programming (SOCP) for optimizing the coordinated transmit beamforming, and develop two efficient designs for updating the reflective beamforming based on the techniques of semi-definite relaxation (SDR) and successive convex approximation (SCA), respectively. Numerical results show that the use of IRS leads to significantly higher SINR values than benchmark schemes without IRS or without proper reflective beamforming optimization; while the developed SCA-based solution outperforms the SDR-based one with lower implementation complexity.
\end{abstract}

\begin{IEEEkeywords}
Intelligent reflecting surface (IRS), multi-cell systems, multiple-input single-output (MISO), coordinated transmit beamforming, reflective beamforming, optimization.
\end{IEEEkeywords}

\newtheorem{definition}{\underline{Definition}}[section]
\newtheorem{fact}{Fact}
\newtheorem{assumption}{Assumption}
\newtheorem{theorem}{\underline{Theorem}}[section]
\newtheorem{lemma}{\underline{Lemma}}[section]
\newtheorem{corollary}{\underline{Corollary}}[section]
\newtheorem{proposition}{\underline{Proposition}}[section]
\newtheorem{example}{\underline{Example}}[section]
\newtheorem{remark}{\underline{Remark}}[section]
\newtheorem{algorithm}{\underline{Algorithm}}[section]
\newcommand{\mv}[1]{\mbox{\boldmath{$ #1 $}}}

\section{Introduction}

To enable emerging Internet of things (IoT) and artificial intelligence (AI) applications, the fifth-generation (5G)-and-beyond cellular networks need to support massive wireless devices with diverse quality of service (QoS) requirements, such as significantly increased spectrum efficiency, ultra-low transmission latency, and  extremely-high communication reliability \cite{5G,6G}. Towards this end, small base stations (BSs) are densely deployed to shorten the distances with cellular subscribers \cite{small_cells}, and device-to-device (D2D) communications are enabled underlying conventional cellular transmissions to create more spectrum reuse opportunities \cite{D2D}. However, the emergence of small BSs and D2D communications in 5G-and-beyond cellular networks also introduces severe co-channel interference among different cells and different D2D links, which needs to be carefully dealt with from technical perspectives. In the literature, various approaches have been proposed to mitigate or even utilize the co-channel interference, some examples including coordinated beamforming \cite{Co_beam_2010,Co_beam_2011,ICIC,Net_MIMO3} and network multiple-input multiple-output (MIMO) \cite{Net_MIMO1,Net_MIMO2,Net_MIMO4}.

Recently, intelligent reflecting surface (IRS) has emerged as a promising technology for beyond-5G cellular networks \cite{IRS_wu,IRS_survey}, which can also be used to tackle the critical co-channel interference issue in a cost-effective manner. IRS is a passive meta-material panel consisting of a large number of reflecting units, each of which can introduce an independent phase shift on radio-frequency (RF) signals to change the signal transmission environment. By jointly controlling these phase shifts, the IRS can form reflective signal beams, such that the reflected signals can be coherently combined with the directly transmitted signals at intended receivers for enhancing the desirable signal strength, or destructively combined at unintended receivers for suppressing the undesirable interference. As the IRS is a passive device with no dedicated power consumption, it is envisioned as a green and cost-effective solution to enhance the spectrum- and energy-efficiency of future cellular networks \cite{IRS_wu,IRS_survey}.

There have been some prior works \cite{IRS_single1,IRS_single2,IRS_multiuser,IRS_NOMA1,IRS_NOMA2,IRS_OFDM,IRS_OFDM_est,IRS_OFDM_protocol} investigating the joint transmit and reflective beamforming design in IRS-aided wireless communication systems. The authors in \cite{IRS_single1,IRS_single2} investigated the received signal-to-noise ratio (SNR)  maximization problem in a point-to-point IRS-aided multiple-input single-output (MISO) communication system, which is solved by using the techniques of semi-definite relaxation (SDR) \cite{IRS_single1} and manifold optimization \cite{IRS_single2}, respectively. Furthermore, \cite{IRS_multiuser} considered the signal-to-interference-plus-noise ratio (SINR)-constrained power minimization problem in IRS-aided multiuser MISO downlink communication systems, in which alternating optimization is employed to update the transmit and reflective beamforming vectors in an alternating manner, and SDR is employed to optimize the reflective beamforming. In addition, prior works also studied other communication setups aided by the IRS such as IRS-aided orthogonal frequency division multiplexing (OFDM) \cite{IRS_OFDM_protocol,IRS_OFDM_est,IRS_OFDM}, non-orthogonal multiple access (NOMA) \cite{IRS_NOMA1,IRS_NOMA2} and simultaneous wireless information and power transfer (SWIPT) systems \cite{SWIPT}. Nevertheless, all the above prior works \cite{IRS_single1,IRS_single2,IRS_multiuser,IRS_NOMA1,IRS_NOMA2,IRS_OFDM,IRS_OFDM_est,IRS_OFDM_protocol,SWIPT}  focused on a single-cell setup. This thus motivates us to use IRSs to facilitate the multi-cell communications in this work. 

%\begin{figure}
%\centering
% \epsfxsize=1\linewidth
%    \includegraphics[width=6.8cm]{IRS_Multicell3.eps}
%\caption{Illustration of an IRS-aided multi-cell MISO system.} \label{fig:Multicell}
%\vspace{-2em}
%\end{figure}

In this paper, we consider an IRS-aided multi-cell MISO system, where an IRS is dedicatedly deployed at the cell boundary to assist the wireless transmission from BSs to users and suppress their inter-cell interference. We assume that there is one multi-antenna BS serving one single-antenna user in each cell. Our objective is to jointly optimize the coordinated transmit beamforming at the multiple BSs and the reflective beamforming at the IRS, to maximize the minimum weighted received SINR at users, subject to the individual maximum transmit power constraints at BSs, and the reflection constraints at the IRS. However, the formulated minimum SINR maximization problem is highly non-convex due to the coupling between the transmit and reflective beamforming vectors. To solve this difficult problem, we propose efficient algorithms based on alternating optimization, in which the transmit and reflective beamforming vectors are optimized in an alternating manner. In particular, under any given reflective beamforming, we obtain the optimal coordinated transmit beamforming via second-order cone programming (SOCP); while under any given coordinated transmit beamforming, we develop two efficient designs to update the reflective beamforming by using the techniques of SDR and successive convex approximation (SCA), respectively. It is observed that the performance of the SDR-based solution generally depends on the randomization procedure, while the SCA-based solution can always converge towards a stationary point. Numerical results show that the use of IRS leads to significant performance gains over benchmark schemes without IRS or without proper reflective beamforming design at the IRS, and the developed SCA-based solution outperforms the SDR-based one with lower implementation complexity.

It is worth noting that there is only one existing work \cite{IRS_multicell} that studied the weighted sum-rate maximization in IRS-aided multi-cell networks by applying the alternating-optimization-based approaches. Nevertheless, this paper is different from \cite{IRS_multicell} in the following two aspects. First, while \cite{IRS_multicell} focsed on the weighted sum-rate maximization, this paper considers a different objective of the min-weighted-SINR maximization. Second, while \cite{IRS_multicell} only optimized the reflection phases at the IRS by considering unit amplitudes, this paper further exploits the optimization of reflection amplitudes to enhance the communication performance.

\section{System Model and Problem Formulation}

%As shown in Fig. \ref{fig:Multicell}, 
We consider an IRS-aided multi-cell MISO system, where an IRS is dedicatedly deployed to assist the multi-cell communication and suppress the inter-cell interference, especially for cell-edge users. Suppose that in each cell there is a BS with $M\ge1$ antennas communicating with a user with one single antenna. Let $\mathcal K \triangleq\{1,\ldots,K\}$ denote the set of BSs or users in the system, and $\mathcal N \triangleq\{1,\ldots,N\}$ denote the set of reflecting units at the IRS. The IRS can adaptively adjust the reflecting phases to form reflective signal beams, such that the reflected signal can be coherently combined with the directly transmitted signal at the intended user or destructively combined at the unintended users.

We consider a quasi-static narrow-band channel model, where the wireless channels remain unchanged within each transmission block of our interest but may change over different blocks. Let $\mv{G}_i\in {\mathbb C}^{N\times M}$ denote the channel matrix from BS $i$ to the IRS, ${\mv{f}_{i}}\in {\mathbb C}^{N\times 1}$ denote the channel vector from the IRS to user $i$, and ${\mv{h}_{i,k}}\in {\mathbb C}^{M\times 1}$ denote that from BS $k$ to user $i$, where $\mathbb{C}^{x\times y}$ denotes the space of $x\times y$ complex matrices.

Let $s_i$ denote the transmitted signal by each BS $i$ and $\mv w_i\in\mathbb C^{M\times1}$ the corresponding transmit beamforming vector. We assume that $s_i$'s are independent and identically distributed (i.i.d.) circularly symmetric complex Gaussian (CSCG) random variables with zero mean and unit variance, i.e., $s_i\!\sim\!\mathcal{CN}(0,1)$. The transmitted signal by each BS $i$ is thus given by $\mv x_i = \mv w_i s_i, \forall i\in\mathcal K.$ Suppose that each BS has a maximum power budget denoted by $P_i$. Then we have $\mathbb{E}(\|\mv x_i\|^2)\!=\!\|\mv w_i\|^2\!\le\!P_i, \forall i\!\in\!\mathcal K$, where $\mathbb{E}(\cdot)$ denotes the stochastic expectation.

As for the reflection at the IRS, let $\theta_n\!\in\![0, 2\pi)$ and $\beta_n\in[0, 1]$ denote the phase shift and the reflection amplitude imposed by the $n$-th reflecting unit on the incident signal, respectively. Accordingly, let $\mv{\Theta}=\mathrm{diag} (\beta_1 e^{j\theta_{1}},\ldots,\beta_N e^{j\theta_{N}} ) $ represent the reflection coefficient matrix at the IRS, where $j\triangleq \sqrt{-1}$, and $\mathrm{diag}(a_1,\ldots, a_N)$ denotes a diagonal matrix with its diagonal elements being $a_1,\ldots, a_N$. Furthermore, let $\mv v=[\beta_1 e^{j\theta_{1}},\ldots,\beta_N e^{j\theta_{N}}]^H$ denote the reflective beamforming vector, where each element $n$, denoted by $v_n$, must satisfy $|v_n|\le 1, \forall n\in\mathcal N$. Here, the superscript $H$ denotes the conjugate transpose of a vector or matrix. As a consequence, we have the combined reflective channel from BS $k$ to user $i$ as $\mv{f}^H_{i}\mv{\Theta}\mv{G}_k = \mv v^H\mv\Phi_{i,k}$, where $\mv\Phi_{i,k} = \mathrm{diag}(\mv{f}^H_{i})\mv{G}_k$. Notice that this transformation separates the reflective beamforming vector $\mv v$ from the reflective channels, which will significantly facilitate our derivation later. By combining the directly transmitted and reflected signals, the signal received at user $i$ is accordingly expressed as
\begin{align}
y_{i}\!=\!(\mv v^H\mv\Phi_{i,i}\!+\!\mv h^H_{i,i})\mv w_i s_i\!+\!\sum\limits_{k\ne i,k\in\mathcal K}( \mv v^H\mv\Phi_{i,k}\!+\!\mv h^H_{i,k})\mv w_k s_k\!+\!n_{i}, \label{fm:1}
\end{align}
where $n_{i}$ denotes the additive white Gaussian noise (AWGN) at the receiver of user $i$ with zero mean and variance $\sigma_i^2$, i.e., $n_{i}\sim\mathcal{CN}(0,\sigma_i^2), \forall i\in\mathcal K$. By treating the interference as noise, the received SINR at user $i$ is given by
\begin{align}
\mathrm{\gamma}_i(\mv v, \{\mv w_i\})\!=\!\frac{|(\mv v^H\mv\Phi_{i,i}+ \mv h^H_{i,i})\mv w_i|^2}{\sum\limits_{k\ne i,k\in\mathcal K}|(\mv v^H\mv\Phi_{i,k}+ \mv h^H_{i,k})\mv w_k|^2+\sigma_i^2}.\label{fm:2}
\end{align}

Our objective is to maximize the users' communication performance in a fair manner. As a result, we consider the max-min fairness problem with the objective of maximizing the minimum weighted SINR of all users, by jointly optimizing the transmit beamforming $\{\mv w_i\}$ at the BSs and the reflective beamforming $\mv v$ at the IRS, subject to the individual transmit power constraints at BSs and the reflection constraints at the IRS. Let $\alpha_i>0$ denote a weight parameter for user $i\in\mathcal K$ characterizing the fairness among the $K$ users, where a larger value of $\alpha_i$ indicates that user $i$ has a higher priority in transmission. Therefore, the minimum SINR maximization problem is formulated as
\begin{align}
\mathtt{(P1)}:&\mathop\mathtt{max}_{\mv v,\{\mv w_i\}}\mathop\mathtt{min}_{i\in\mathcal K}~\frac{\mathrm{\gamma}_i(\mv v, \{\mv w_i\})}{\alpha_i} \label{Problem:ori:1} \\
&~~~{\mathtt{s.t.}}~~\|\mv w_i\|^2 \le P_i, \forall i\in\mathcal K \label{Problem:ori:2} \\
&~~~~~~~~~~|v_n|\le 1, \forall n\in\mathcal N. \label{Problem:ori:3}
\end{align}
To facilitate the derivation, we first introduce an auxiliary variable $t$ and reformulate problem (P1) as the following equivalent problem:
\begin{align}
&\mathtt{(P1.1)}:\mathop\mathtt{max}_{\mv v,\{\mv w_i\},t} t \nonumber \\
&~~~~~~~~~~~{\mathtt{s.t.}}
~~~\mathrm{\gamma}_i(\mv v, \{\mv w_i\})\ge\alpha_i t, \forall i\in\mathcal K \label{Problem:ori_epi:1} \\
&~~~~~~~~~~~~~~~~~~~(\ref{Problem:ori:2})~\text{and}~(\ref{Problem:ori:3}). \nonumber
\end{align}
Notice that problem (P1.1) or (P1) is difficult to be optimally solved due to the coupling between the transmit beamforming $\{\mv w_i\}$ and the reflective beamforming $\mv v$ at the SINR terms. To tackle this difficulty, we propose alternating-optimization-based algorithms to solve problem (P1.1) or (P1), in which the transmit beamforming vector $\{\mv w_i\}$ and the reflective beamforming vector $\mv v$ are optimized in an alternating manner, with the other being fixed. In particular, the alternating-optimization-based algorithms are implemented in an iterative manner. For notational convenience, suppose that at each iteration $l \ge 0$, the obtained beamforming vectors are denoted by $\mv v^{(l)}$ and $\{\mv w^{(l)}_i\}$, where $\mv v^{(0)}$ and $\{\mv w^{(0)}_i\}$ denote the initial beamforming vectors. In Sections \ref{sec:III} and \ref{sec:IV}, we present efficient approaches for updating $\{\mv w_i\}$ and $\mv v$, respectively.

\section{Coordinated Transmit Beamforming Optimization}\label{sec:III}

In this section, we optimize the coordinated transmit beamforming $\{\mv w_i\}$ under any given reflective beamforming $\mv v$. For notational convenience, we define $\mv a_{i,k}=\mv\Phi_{i,k}^H\mv v + \mv h_{i,k}$ as the effective or combined channel from BS $k\in\mathcal K$ to user $i\in\mathcal K$. Accordingly, the coordinated transmit beamforming optimization problem becomes
\begin{align}
\mathrm{(P2):}&\mathop\mathtt{max}_{\{\mv w_i\},t} ~ t \nonumber \\
&~~~{\mathtt{s.t.}}
~\frac{|\mv a^H_{i,i}\mv w_i|^2}{\sum\limits_{k\ne i,k\in\mathcal K}|\mv a^H_{i,k}\mv w_k|^2+\sigma_i^2}\ge\alpha_i t, \forall i\in\mathcal K, \label{Problem:given_phase:1} \\
&~~~~~~~~~(\ref{Problem:ori:2}). \nonumber
\end{align}
It is observed that problem (P2) is still not a convex optimization problem. To tackle this issue, we introduce the following feasibility problem (P2.1), which is obtained based on problem (P2) by fixing $t$.
\begin{align}
\mathrm{(P2.1):}&\mathop\mathtt{find}~ \{\mv w_i\} \nonumber \\
&~~~{\mathtt{s.t.}}
~(\ref{Problem:ori:2})~\mathrm{and}~(\ref{Problem:given_phase:1}). \nonumber
\end{align}
In particular, suppose that the optimal solution of $t$ to problem (P2) is given by $t^\star$. It is thus clear that if problem (P2.1) is feasible under any given $t$, then we have $t \le t^\star$; while if (P2.1) is infeasible, then it follows that $t > t^\star$. Therefore, problem (P2) can be equivalently solved by checking the feasibility of problem (P2.1) under any given $t > 0$, together with a bisection search over $t > 0$.

Therefore, to solve problem (P2), we only need to solve problem (P2.1) under any fixed $t > 0$, by using SOCP as follows \cite{SOCP}. Towards this end, we notice that the SINR constraints in (\ref{Problem:given_phase:1}) can be reformulated as
\begin{align}
\big(1+\frac{1}{\alpha_i t}\big)|\mv a^H_{i,i}\mv w_i|^2 \ge \sum\limits_{k\in\mathcal K}|\mv a^H_{i,k}\mv w_k|^2+\sigma_i^2, \forall i\in\mathcal K. \label{tf_socp:1}
\end{align}
Based on (\ref{tf_socp:1}), it is evident that if $\{\mv w_i\}$ is a feasible solution to problem (P2.1), then any phase rotation of $\{\mv w_i\}$ will still be feasible. Without loss of optimality, we choose the solution of $\{\mv w_i\}$ such that $\mv a^H_{i,i}\mv w_i$ becomes a non-negative value for any user $k\in  \mathcal K$. As a result, we have the following constraints:
\begin{align}
\mv a^H_{i,i}\mv w_i \ge 0, \forall i\in\mathcal K,\label{tf_socp:2}
\end{align}
where $\mv a^H_{i,i}\mv w_i$ has a non-negative real part and a zero imaginary part, i.e., $\mathrm{Re}(\mv a_{i,i}^H \mv w_i) \ge 0$ and $\mathrm{Im}(\mv a^H_{i,i}\mv w_i)=0$, with $\mathrm{Re}(x)$ and $\mathrm{Im}(x)$ denoting the real and imaginary parts of a complex number $x$. Accordingly, (\ref{tf_socp:1}) can be further re-expressed as
\begin{align}
\sqrt{1+\frac{1}{\alpha_i t}}\mv a^H_{i,i}\mv w_i \ge \begin{Vmatrix}
\mv A^H \mv e_i \\
\sigma_i
\end{Vmatrix}_2, \forall i\in\mathcal K, \label{tf_socp:3}
\end{align}
where $\mv A\in\mathbb C^{K\times K}$ denotes a matrix with the element in its $i$-th row and $j$-th column being $\mv a^H_{i,j}\mv w_j$, $\mv e_i\in\mathbb C^{K\times 1}$ denotes a vector with the $i$-th element being one and others being zero, and $\|\cdot\|_2$ denotes the Euclidean norm of a vector. Therefore, problem (P2.1) is reformulated as the following equivalent form:
\begin{align}
\mathrm{(P2.2):}\mathop\mathtt{find}~& \{\mv w_i\} \nonumber \\
{\mathtt{s.t.}}
~& (\ref{Problem:ori:2}),(\ref{tf_socp:2}),~\mathrm{and}~(\ref{tf_socp:3}). \nonumber
\end{align}
Problem (P2.1) is an SOCP problem that can be optimally solved by standard convex optimization solvers such as CVX \cite{CVX}. Therefore, the optimal coordinated transmit beamforming solution to problem (P2) is finally obtained.

\section{Reflectve Beamforming Optimization}\label{sec:IV}
In this section, we optimize the reflective beamforming vector $\mv v$ under given transmit beamforming $\{\mv w_i\}$. For notational convenience, we define $\mv c_{i,k} = \mv\Phi_{i,k}\mv w_k$ and $d_{i,k} = \mv h^H_{i,k}\mv w_k, \forall i, k\in\mathcal K$. Then, we have
\begin{align}
|(\mv v^H\mv{\Phi}_{i,k}\!+\!\mv h^H_{i,k})\mv w_k|^2\!=\!\mv v^H\mv C_{i,k}\mv v\!+\!2\mathrm{Re}\{\mv v^H\mv u_{i,k}\}\!+\!|d_{i,k}|^2, \label{fm:3}
\end{align}
where $\mv C_{i,k}=\mv c_{i,k}\mv c^H_{i,k}$ and $\mv u_{i,k}=\mv c_{i,k}d_{i,k}^H$.
Accordingly, the reflective beamforming optimization problem is given by
\begin{align}
&\mathtt{(P3):}\mathop\mathtt{max}_{\mv v,t}~t \nonumber \\
&~~~~~~~~{\mathtt{s.t.}}
~\frac{\mv v^H\mv C_{i,i}\mv v\!+\!2\mathrm{Re}\{\mv v^H\mv u_{i,i}\}\!+\!|d_{i,i}|^2}{\sum\limits_{k\ne i,k\in\mathcal K}\mv v^H\mv C_{i,k}\mv v\!+\!2\mathrm{Re}\{\mv v^H\mv u_{i,k}\}\!+\!|d_{i,k}|^2\!+\!\sigma_i^2}\nonumber \\
&~~~~~~~~~~~~~~~~~~~~~~~~\ge\alpha_i t, \forall i\in\mathcal K, \label{Problem:given_beam:1} \\
&~~~~~~~~~~~~~~(\ref{Problem:ori:3}). \nonumber
\end{align}
Notice that problem (P3) is also a non-convex optimization problem. In the following, we propose two solutions by leveraging the SDR and SCA techniques, respectively.

\subsection{SDR-based Solution to Problem (P3)}
In this subsection, we use the well-established SDR technique to solve problem (P3). This is motivated by the wide application of SDR in reflective beamforming optimization (see, e.g., \cite{IRS_multiuser}). Towards this end, we first define $|(\mv{v}^H\mv{\Phi}_{i,k} + \mv h^H_{i,k})\mv w_k|^2 = \bar{\mv v}^{\rm{H}}{\mv R}_{i,k}\bar{\mv v} + |d_{i,k}|^2$, where
\begin{align*}
{\mv R}_{i,k} = \begin{bmatrix}
\mv C_{i,k} & \mv u_{i,k} \\
\mv u^H_{i,k} & 0
\end{bmatrix} ~\mathrm{and} ~\bar{\mv v} = \begin{bmatrix}
\mv{v} \\ 1 \end{bmatrix}.
\end{align*}
Accordingly, problem (P3) is re-expressed as
\begin{align}
&\mathtt{(P3.1):}\mathop\mathtt{max}_{\bar{\mv v},t}~t \nonumber \\
&~~~{\mathtt{s.t.}}
~\frac{\bar{\mv v}^{H}{\mv R}_{i,i}\bar{\mv v} + |d_{i,i}|^2}{\sum\limits_{k\ne i,k\in\mathcal K}\bar{\mv v}^{H}{\mv R}_{i,k}\bar{\mv v} + |d_{i,k}|^2 +\sigma_i^2}\ge\alpha_i t, \forall i\in\mathcal K \label{Problem:given_beam:2} \\
&~~~~~~~~~|\bar v_n|\le 1,|\bar v_{N+1}| = 1, \forall n\in\mathcal N \label{Problem:given_beam:3}
\end{align}
Furthermore, we define $\mv V= \bar{\mv v}\bar{\mv v}^H$ with $\mv V$ being positive semi-definite (i.e., $\mv V \succeq 0$) and $\mathrm{rank}(\mv V)\le 1$. Then problem (P3.1) or (P3) is further reformulated as the following equivalent form:

\begin{align}
&\mathtt{(P3.2):}\mathop\mathtt{max}_{\mv V,t} ~ t \nonumber \\
&~~{\mathtt{s.t.}}
~\frac{\mathrm{Tr}({\mv R_{i,i}}\mv V)\!+\!|d_{i,i}|^2}{\sum\limits_{k\ne i,k\in\mathcal K}\mathrm{Tr}({\mv R_{i,k}}\mv V)\!+\!|d_{i,k}|^2\!+\!\sigma^2_i}\ge\alpha_i t, \forall i\in\mathcal K \label{Problem:given_beam_SDR:1}\\
&~~~~~~ V_{n,n}\le 1, V_{N+1,N+1}=1, \forall n\in\mathcal N \label{Problem:given_beam_SDR:2}\\
&~~~~~~\mv V\succeq 0 \label{Problem:given_beam_SDR:3}\\
&~~~~~~\mathrm{rank}(\mv V)\le 1, \label{Problem:given_beam_SDR:5}
\end{align}
where $V_{m,n}$ denotes the element in the $m$-th row and $n$-th column of the matrix $\mv V$, and $\mathrm{Tr}(\mv A)$ denotes the trace of matrix $\mv A$. However, problem (P3.2) is still challenging to be optimally solved due to the non-convex rank-one constraint in (\ref{Problem:given_beam_SDR:5}). Motivated by the idea of SDR, we relax this constraint, and obtain a relaxed version of (P3.2) as
\begin{align}
&\mathtt{(P3.3):}\mathop\mathtt{max}_{\mv V,t} ~ t \nonumber \\
&~~~~~~~~~~{\mathtt{s.t.}}~(\ref{Problem:given_beam_SDR:1}),(\ref{Problem:given_beam_SDR:2}),~\mathrm{and}~(\ref{Problem:given_beam_SDR:3}).\nonumber
\end{align}
Although problem (P3.3) is non-convex, it can be shown, similarly as for problem (P2.1), that (P3.3) can be solved equivalently by solving the following feasibility problem (P3.4) together with a bisection search over $t$.
\begin{align}
&\mathtt{(P3.4):}\mathop\mathtt{find}~ \mv V \nonumber \\
&~~~~~~~~~~~{\mathtt{s.t.}}
~\mathrm{Tr}({\mv R_{i,i}}\mv V)\!+\!|d_{i,i}|^2\ge \nonumber \\
&~~~~~~~~~~~\alpha_i t(\sum\limits_{k\ne i,k\in\mathcal K}\mathrm{Tr}({\mv R_{i,k}}V)\!+\!|d_{i,k}|^2\!+\!\sigma^2_i), \forall i\in\mathcal K \label{Problem:given_beam_f:1}\\
&~~~~~~~~~~~(\ref{Problem:given_beam_SDR:2})~\mathrm{and}~(\ref{Problem:given_beam_SDR:3}).\nonumber
\end{align}
Notice that problem (P3.4) is a convex semi-definite program (SDP) and thus can be solved optimally by using CVX \cite{CVX}. As a result, we have obtained the optimal solution to problem (P3.3), denoted by $\mv V^\star$ and $t^\star$.

Now, it remains to reconstruct the solution to problem (P3.2) or equivalently (P3.1)/(P3) based on $\mv V^\star$ and $t^\star$. In particular, if rank$(\mv V^\star)\le 1$, then $\mv V^\star$ and $t^\star$ are also the optimal solution to problem (P3.2). In this case, we have $\mv V^\star=\bar{\mv v}^\star \bar{\mv v}^{\star H}$, where $\bar{\mv v}^\star$ becomes the optimal solution to problem (P3.1). However, if rank$(\mv V^\star)>1$, then the following Gaussian randomization procedure \cite{Random} needs to be further adopted to produce a high-quality rank-one solution to problem (P3.2) and (P3.1). Specifically, suppose that the eigenvalue decomposition of $\mv V^\star$ is $\mv V^\star=\mv U\mv \Sigma\mv U^H$. Then, we set $\tilde{\mv v} = \mv U\mv\Sigma^{\frac{1}{2}}\mv r$, where $\mv r$ corresponds to a CSCG random vector with zero mean and covariance matrix $\mv I$, i.e., $\mv r\sim\mathcal{CN}(0,\mv I)$. Accordingly, we construct a feasible solution $\bar{\mv v}$ to problem (P3.1) as $\bar v_n= e^{j\mathrm{arg}(\tilde{v}_n/\tilde{v}_{N+1})}$, where $\bar v_n$ and $\tilde v_n$ denote the $n$-th element of vector $\bar{\mv v}$ and $\tilde{\mv v}$, respectively, and $\mathrm{arg}(x)$ denotes the phase of a complex number $x$. To guarantee the performance, the randomization process needs to be implemented multiple times and the best solution among them is selected as the obtained solution to problem (P3.1), denoted by ${\bar{\mv v}}^\star$. In this case, the obtained solution to problem (P3.2) is ${\bar{\mv v}}^\star {\bar{\mv v}}^{\star H}$. Based on the solution of ${\bar{\mv v}}^\star$ to problem (P3.1), we can accordingly obtain the solution of (P3) as $\mv v^\star$. Therefore, the SDR-based algorithm for solving problem (P3) is complete.

By alternately implementing the SDR-based solution to (P3) and the SOCP-based solution to (P2), we can obtain an efficient solution to the original problem (P1). We refer to this algorithm as alternating optimization with SDR. In summary, the algorithm of alternating optimization with SDR is presented as Algorithm 1.

\begin{remark}
It is worth noticing that the performance of the SDR-based solution to problem (P3) critically depends on the performance of the Gaussian randomization when the rank of the obtained $\mv V^\star$ is larger than one. This results in the following drawbacks for the algorithm of alternating optimization with SDR for solving (P1). On one hand, the SDR may introduce increased implementation complexity, which is due to the fact that solving the SDR is generally time-consuming (especially when the dimension of the matrix becomes large) and a large number of randomizations are generally needed in order to get a better solution. On the other hand, the alternating optimization with SDR may lead to compromised performance, as alternating optimization may terminate if the SDR leads to a highly suboptimal solution due to the uncertainty in randomizations. Therefore, this motivates us to further develop an alternative algorithm with performance guarantee.
\end{remark}

%\makeatletter\def\@captype{table}\makeatother
\begin{table}[htp]
\begin{center}
%\caption{}
\hrule
\vspace{0.1cm}\textbf{Algorithm 1}: Alternating optimization with SDR  \vspace{0.1cm}
\hrule \vspace{0.0cm}
\begin{itemize}
\item[1:] Initialize: $l=0$, $\mv v^{(0)}$ and accuracy threshold $\epsilon > 0$.
\item[2:] {$\mv{\mathrm{Repeat}}$:}
\item[3:] ~~~$l=l+1$;
\item[4:] ~~~Under given $\mv v^{(l-1)}$, solve problem (P2) to obtain $\{\mv w_i^\star\}$, and set $\mv w_i^{(l)}=\mv w_i^\star, \forall i\in \mathcal K$;
\item[5:] ~~~Under given $\{\mv w_i^{(l)}\}$, solve problem (P3) to obtain $\mv v^\star$, and set $\mv v^{(l)}=\mv v^\star$;
\item[6:] $\mv{\mathrm{Until}}$ the increase of the objective function in (P1) is smaller than  $\epsilon$.
\end{itemize}
\vspace{0.0cm} \hrule \label{Table:1}
\end{center}
\end{table}
\vspace{-0.5cm}
\subsection{SCA-based Design for Updating Reflective Beamforming}
To overcome the above drawbacks of the SDR-based solution, in this subsection, we propose an efficient design for updating the reflective beamforming vector $\mv v$, by applying the SCA technique. Recall that the update of $\mv v$ in problem (P3) is implemented iteratively in the alternating-optimization-based algorithm for solving the original problem (P1). Therefore, instead of directly solving (P3), in the SCA-based design we aim to find an updated $\mv v$ to increase the users' minimum SINR. Towards this end, we consider a particular iteration $l\ge 1$, the local point of $\mv v$ as ${\mv v}^{(l-1)}$, which corresponds to the obtained $\mv v$ in the previous iteration. Under given $\{\mv w_i^{(l)}\}$ together with ${\mv v}^{(l-1)}$, we denote the achieved minimum SINR at users as $ t^{(l)} = \min_{i\in\mathcal K} \gamma_i({\mv v}^{(l-1)}, \{\mv w_i^{(l)}\})$. In the following, we explain how to update $\mv v$ to increase the minimum SINR at users based on SCA.

For notational convenience, we first define an auxiliary function for user $i\in \mathcal K$ as
\begin{align}
&\mathcal{F}_i(\mv v, \{\mv w_i\}, t) \nonumber\\
&=\alpha_i t  [\sum\limits_{k\ne i,k\in\mathcal K} (\mv v^H\mv C_{i,k}\mv v\!+\!2\mathrm{Re}\{\mv v^H\mv u_{i,k}\}\!+\!|d_{i,k}|^2 )+\sigma^2_i ] \nonumber \\
&~~~~-\! (\mv v^H\mv C_{i,i}\mv v\!+\!2\mathrm{Re}\{\mv v^H\mv u_{i,i}\}\!+\!|d_{i,i}|^2 ),\label{function}
\end{align}
where $\mv C_{i,k}$, $\mv u_{i,k}$, and $d_{i,k}$, $i, k\in\mathcal K$ are defined at the beginning of Section IV. Note that after the update of $\{\mv w_i\}$ at each iteration $l$, it must hold that $\min_{i\in\mathcal K} \mathcal F_i(\mv v^{(l-1)},\{\mv w_i^{(l)}\}, t^{(l)})\!=\!0$. Accordingly, we update the reflective beamforming vector $\mv v$ at the IRS by solving the following problem:
\begin{align}
\mathtt{(P4):}&\mathop\mathtt{min}_{\mv v} ~\mathop\mathtt{max}_{i\in \mathcal K} ~\mathcal{F}_i (\mv v, \{\mv w_i^{(l)}\}, t^{(l)} ) \label{Problem:given_beam2:1} \\
&~{\mathtt{s.t.}}
~(\ref{Problem:ori:3}). \nonumber
\end{align}
As $\mv v^{(l-1)}$ is a feasible solution to problem (P4) leading to an objective value of zero, the optimal solution to problem (P4) should be non-positive. Suppose that the obtained solution to (P4) as $\mv v^{(l)}$. If $\min_{i\in\mathcal K}\mathcal F(\mv v^{(l)}, \{\mv w_i^{(l)}\}, t^{(l)}) < 0$, then it can be easily shown that $\min_{i\in\mathcal K} \gamma_i(\mv v^{(l)}, \{\mv w_i^{(l)}\}) >\min_{i\in\mathcal K} \gamma_i(\mv v^{(l-1)}, \{\mv w_i^{(l)}\})$, i.e., the minimum SINR is increased. Therefore, we focus on solving problem (P4) next.

Problem (P4) is still non-convex as the objective function is non-convex with respect to $\mv v$. To address this issue, we apply the SCA technique to approximate the second convex term in the right-hand-side of (\ref{function}) by its first-order Taylor expansion. Note that a convex function is lower bounded by its first-order Taylor expansion at any given point. Thus, at the local point of $\mv v^{(l-1)}$, we have
\begin{align}
&\mathcal{F}_i(\mv v, \{\mv w_i^{(l)}\}, t^{(l)})\le \nonumber \\
&~~~\alpha_i t^{(l)}[\sum\limits_{k\ne i,k\in\mathcal K} (\mv v^H\mv C_{i,k}\mv v\!+\! 2\mathrm{Re}\{\mv v^H\mv u_{i,k}\}\!+\!|d_{i,k}|^2 )+\sigma^2_i] \nonumber \\
&~~~-\! ({\mv v^{(l-1)}}^H\mv C_{i,i}{\mv v^{(l-1)}}\!+\!2\mathrm{Re}\{{\mv v^{(l-1)}}^H\mv u_{i,i}\}\!+\!|d_{i,i}|^2 ) \nonumber \\
&~~~-\!2 (\mv C^H_{i,i}{\mv v^{(l-1)}}\!+\!\mv u_{i,i} )^H (\mv v\!-\!\mv v^{(l-1)} )\nonumber \\
&~~~\triangleq\mathcal{F}^{\mathtt{up}}_i (\mv v, \{\mv w_i^{(l)}\}, t^{(l)},\mv v^{(l-1)} ).\label{upper_bound}
\end{align}
By introducing an auxiliary variable $z$ and replacing $\mathcal{F}_i (\mv v, \{\mv w_i^{(l)}\}, t^{(l)} )$ by $\mathcal{F}^{\mathtt{up}}_i (\mv v, \{\mv w_i^{(l)}\}, t^{(l)},\mv v^{(l-1)} )$, problem (P4) is approximated as the following problem:
\begin{align}
\mathtt{(P4.1):}\mathop\mathtt{min}_{\mv v, z} ~& z \nonumber \\
{\mathtt{s.t.}}
~&\mathcal{F}^{\mathtt{up}}_i (\mv v, \{\mv w_i^{(l)}\}, t^{(l)},\mv v^{(l-1)} )\le z, \forall i\in\mathcal K, \label{Problem:given_beam_SCA:1} \\
~&(\ref{Problem:ori:3}). \nonumber
\end{align}
Problem (P4.1) is a convex problem that can be solved optimally by CVX \cite{CVX}. Suppose that the optimal solution to problem (P4.1) is denoted as $\mv v^{\star\star}$ and $t^{\star\star}$. By substituting $\mv v^{\star\star}$ into $\mathcal{F}_i (\mv v, \{\mv w_i^{(l)}\}, t^{(l)} )$, it is evident that as any feasible solution problem (P4.1) is also feasible for problem (P4), the obtained value of (P4.1) by $\mv v^{\star\star}$ is smaller than that of (P4). Therefore, $\mv v^{\star\star}$ leads to an increased SINR value. Therefore, we can directly update $\mv v$ as $\mv v^{\star\star}$, i.e., $\mv v^{(l)} = \mv v^{\star\star}$. In summary, the alternating optimization with SCA is presented as Algorithm 2.

\vspace{-0.2cm}
\begin{table}[htp]
\begin{center}
%\caption{}
\hrule
\vspace{0.1cm} \textbf{Algorithm 2}: Alternating optimization with SCA  \vspace{0.1cm}
\hrule \vspace{0.0cm}
\begin{itemize}
\item[1:] Initialize: $l=0$, $\mv v^{(0)}$ and accuracy threshold $\epsilon > 0$.
\item[2:] {$\mv{\mathrm{Repeat}}$:}
\item[3:] ~~~$l=l+1$;
\item[4:] ~~~Under given $\mv v^{(l-1)}$, solve problem (P2) to obtain $\{\mv w_i^\star\}$ and $t^\star$, and set $\mv w_i^{(l)}=\mv w_i^\star, \forall i\in \mathcal K,$ and $t^{(l)}=t^\star$;
\item[5:] ~~~Under given $\{\mv w_i^{(l)}\}$, $t^{(l)}$, and $\mv v^{(l-1)}$, solve problem (P4.1) to obtain $\mv v^{\star\star}$, and set $\mv v^{(l)}=\mv v^{\star\star}$;
\item[6:] $\mv{\mathrm{Until}}$ the increase of the objective function in (P1) is smaller than  $\epsilon$.
\end{itemize}
 \hrule\label{Table:2}
\end{center}
\end{table}
\vspace{-0.4cm}
It is worth noting that for the algorithm of alternating optimization with SCA, it is ensured that after each update of $\mv v$ by SCA, the minimum SINR among all users is always non-decreasing. Therefore, the objective value of (P1) is ensured to be non-decreasing at each iteration. As a result, this algorithm will converge towards a stationary solution to problem (P1).

\section{Numerical Results}

In this section, we provide numerical results to validate the performance of the proposed alternating-optimization-based algorithms in the IRS-aided multi-cell MISO system. In the simulation, there are $K=3$ BSs located at $(-100~\text{m},0)$, $(100~\text{m},0)$ and $(0,100~\text{m})$, respectively, each of which is equipped with $M=2$ antennas. We consider a scenario with symmetrically distributed users unless otherwise stated, where the three users are located at $(-d_\text{user},0)$, $(d_\text{user},0)$ and $(0,d_\text{user})$, with $d_\text{user}\!=\!5~\text{m}$. An IRS with $N\!=\!20$ reflecting units is deployed at $(0,-d_\text{IRS})$, with $d_\text{IRS} = 10~\text{m}$. Furthermore, we set the maximum transmit power at all BSs to be identical, i.e., $P_i=P_\text{max},\forall i\in\mathcal K$, and we are interested in the minimum SINR at users by setting $\alpha_i=1,\forall i\in\mathcal K$. In addition, we consider the distance-dependent path loss model as
\begin{align}
P_L=C_0 (\frac{d}{d_0} )^{-\alpha}, \label{fm:16}
\end{align}
where $C_0=-30~$dB denotes the path loss at the reference distance of $d_0=1~$m, $\alpha$ denotes the path loss exponent, $d$ denotes the distance between the transmitter and receiver. For  the BS-user, BS-IRS and IRS-user links, we set the path-loss exponents $\alpha$ to be $3.6$, $2$, and $2.5$, respectively. Furthermore, we consider line-of-sight (LOS) channels from BSs to the IRS, and Rayleigh fading for the BS-user and IRS-user links. The noise power at each user $i$ is set as $\sigma^2_i=-80~{\text{dBm}}, \forall i\in\mathcal K$.

\begin{figure*}[htbp]
	\centering
	\subfigure{
		\begin{minipage}[t]{0.31\linewidth}
			\centering\setcounter{figure}{0} 
			\includegraphics[width=6.2cm]{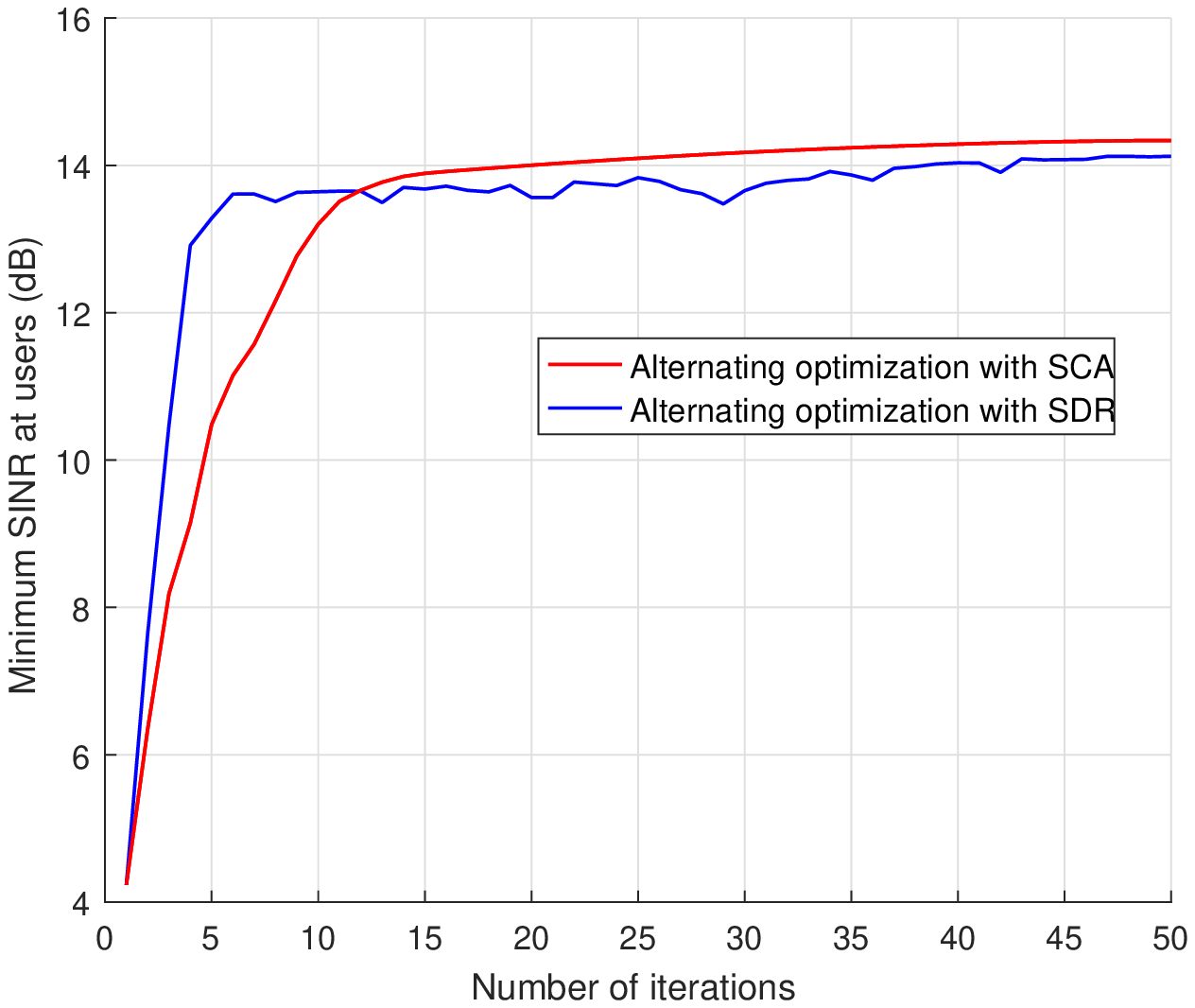}
			\caption{Convergence behavior of the two proposed alternating-optimization-based algorithms.}\label{fig:Iteration}
		\end{minipage}
	}
	\subfigure{
		\begin{minipage}[t]{0.31\linewidth}
			\centering
			\includegraphics[width=6.2cm]{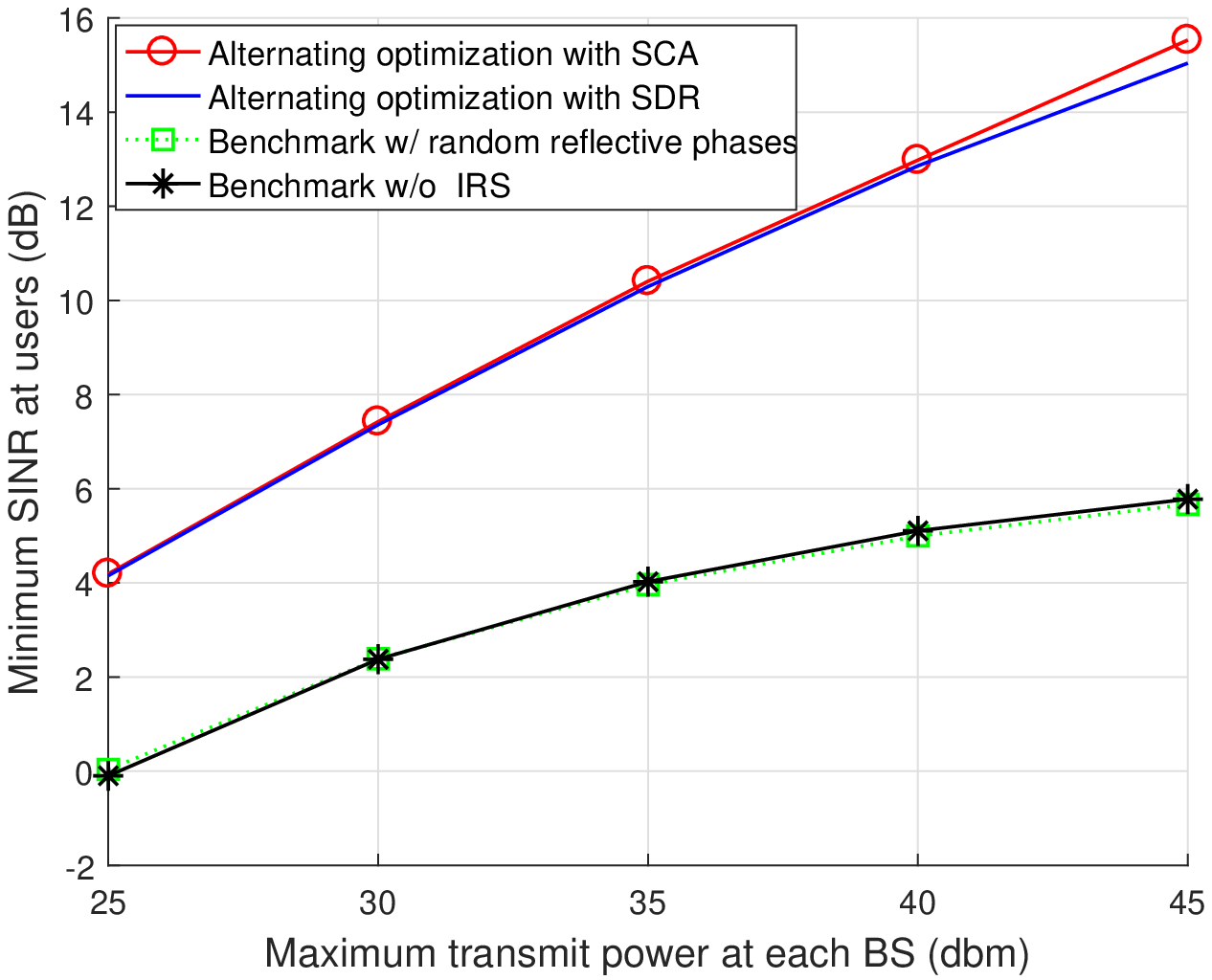}
			\caption{The minimum SINR at users versus the maximum transmit power $P_{\text{max}}$ at each BS, in the scenario with symmetrically distributed users.}\label{fig:Power}
		\end{minipage}
	}
	\subfigure{
		\begin{minipage}[t]{0.31\linewidth}
			\centering
			\includegraphics[width=6.2cm]{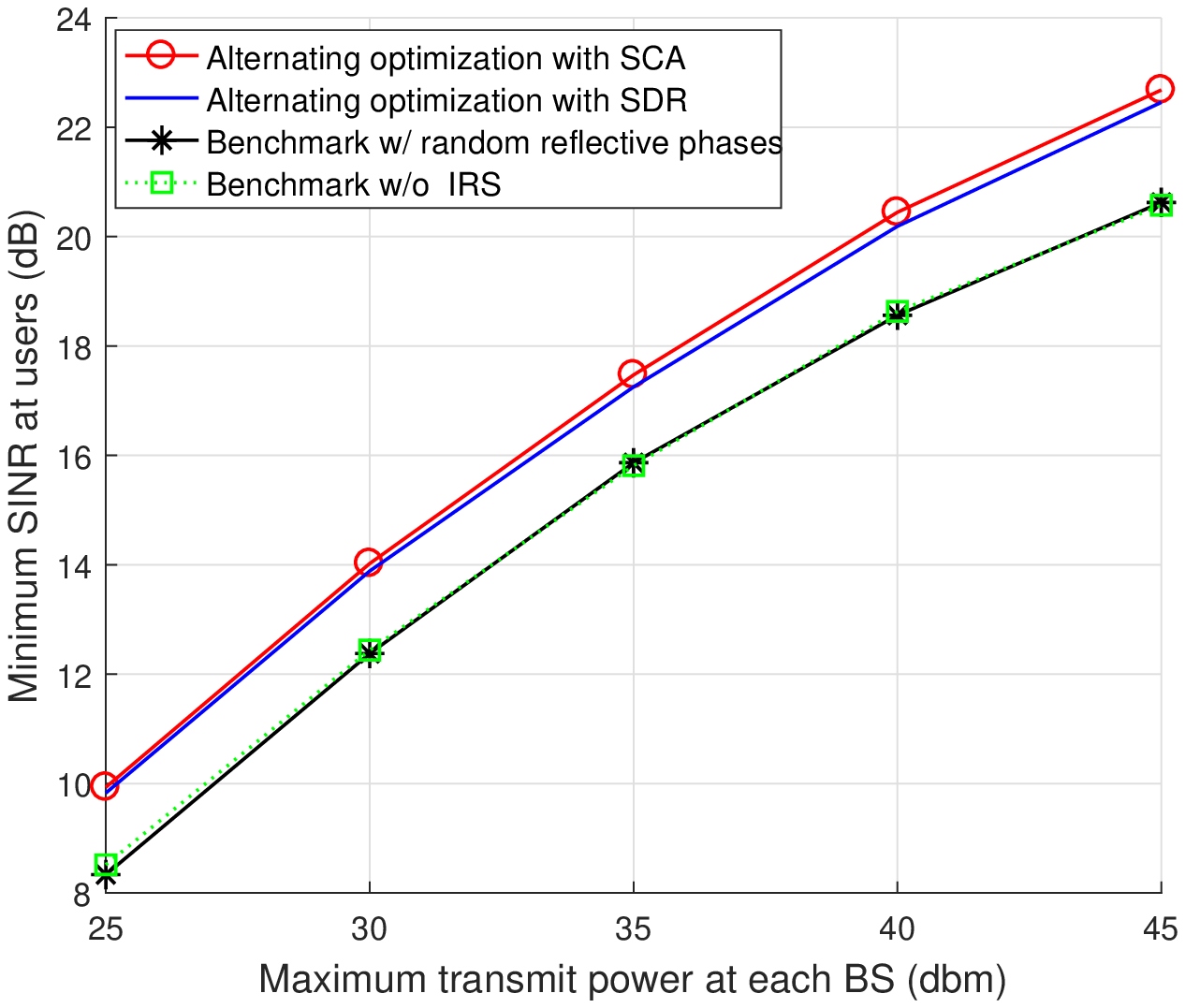}
			\caption{The minimum SINR at users versus the maximum transmit power $P_{\text{max}}$ at each BS,  in the scenario with randomly distributed users.}\label{fig:Random}
		\end{minipage}
	}
	\vspace{-0.2cm}
	\centering
	\vspace{-0.5cm}
\end{figure*}

%\begin{figure}[htbp]
%\vspace{-1em}
%\centering
% \epsfxsize=1\linewidth
%    \includegraphics[width=6cm]{Iteration.eps}
%\caption{Convergence behavior of the two proposed alternating-optimization-based algorithms.} \label{fig:Iteration}
%\vspace{-1em}
%\end{figure}

First, Fig. \ref{fig:Iteration} shows the convergence behaviour of the two alternating-optimization-based algorithms, where $P_\text{max} = 35~$dBm. It is observed that the alternating optimization with SCA leads to monotonically increasing SINR values over iterations, thus converging towards a stationary solution; while the alternating optimization with SDR results in fluctuated SINR values due to the randomization process. Furthermore, it is also observed that running on a computer with E5-2667v4 CPU and 32G memory, the average run time of the alternating optimization with SDR is $895.1628$ seconds, while that of the alternating optimization with SCA is $432.7232$ seconds. This shows the advantage of SCA again in terms of the implementation complexity.

Next, we evaluate the performance of the proposed two alternating-optimization-based algorithms, as compared with the following two benchmark schemes.
\begin{itemize}
\item {\bf Benchmark scheme with random reflective phases}: We set the phase shift $\theta_n$ of each unit $n$ at the IRS as a random value uniformly distributed in $[0, 2\pi)$, and set $\beta = 1$. Under such given reflective beamforming, we solve problem (P2) to obtain the corresponding coordinated transmit beamforming.
\item {\bf Benchmark scheme without IRS}: Without IRS deployed, we only need to optimize the coordinated transmit beamforming by solving problem (P2), in which $\{\mv a_{i,k}\}$ is replaced as $\{\mv h_{i,k}\}$.
\end{itemize}

%\begin{figure}[htbp]
%	\vspace{-1em}
%\centering
% \epsfxsize=1\linewidth
%    \includegraphics[width=6cm]{Power.eps}
%\caption{The minimum SINR at users versus the maximum transmit power $P_{\text{max}}$ at each BS, in the scenario with symmetrically distributed users. } \label{fig:Power}
%\vspace{-1em}
%\end{figure}

Fig. \ref{fig:Power} shows the minimum SINR at users versus the maximum transmit power $P_\text{max}$ at each BS, in which the results are averaged over $100$ random channel realizations. First, it is observed that the two proposed alternating-optimization-based algorithms considerably outperform the two benchmark schemes, and the performance gains become more significant when $P_\text{max}$ gets large. This shows the benefit of IRS in both signal enhancement and interference suppression, especially when the interference (or transmit power) becomes strong. Next, the alternating optimization with SCA is observed to lead to higher minimum SINR values than that with SDR. This is consistent with the observation in Fig. \ref{fig:Iteration}. Furthermore, the benchmark scheme with random reflective phases is observed to have a similar performance as that without IRS. This shows that the benefit of IRS can only be gained via proper reflective beamforming optimization.

%\begin{figure}[htbp]
%	\vspace{-1em}
%\centering
% \epsfxsize=1\linewidth
%    \includegraphics[width=6cm]{Random_Location.eps}
%\caption{The minimum SINR at users versus the maximum transmit power $P_{\text{max}}$ at each BS,  in the scenario with randomly distributed users. } \label{fig:Random}
%\vspace{-1em}
%\end{figure}

To further reveal the practical performance, Fig. \ref{fig:Random} shows the minimum SINR at users versus $P_\text{max}$, in another scenario with the three users uniformly distributed within a triangle area whose vertexes correspond to the three BSs. The results in Fig. \ref{fig:Random} are obtained by averaging over $100$ random user realizations. Similar observations are made in Fig. \ref{fig:Random} as in Fig. \ref{fig:Power}. Nevertheless, the performance gain brought by IRS becomes less significant in this scenario with randomly distributed users, as the users are likely to be located at the cell center, such that the direct BS-user links become strong but the IRS-user links become weak.

\section{Conclusion}
In this paper, we investigated the IRS-aided multi-cell MISO system, with the objective of maximizing the minimum weighted SINR at all users by jointly optimizing the coordinated transmit beamforming at BSs and reflective beamforming at the IRS, subject to individual power constraints at BSs. We proposed efficient alternating-optimization-based algorithms to update the transmit and reflective beamforming vectors in an alternating manner. In particular, we used the SOCP to optimize the transmit beamforming, and proposed two designs based on SDR and SCA for updating the reflective beamforming. Numerical results demonstrated that the dedicatedly deployed IRS considerably improves the performance of the multi-cell MISO system by not only enhancing the received signal strength but also suppressing the inter-cell interference, especially for cell-edge users. It was also shown that the SCA-based design is an efficient algorithm for optimizing the reflective beamforming at the IRS with guaranteed convergence, which outperforms the conventionally adopted SDR-based reflective beamforming optimization.


\begin{thebibliography}{}

\bibitem{5G} J. Andrews, S. Buzzi, W. Choi, S. V. Hanly, A. Lozano, A. Soong, and J. C. Zhang, ``What will 5G be?'' {\it IEEE J. Sel. Areas Commun.}, vol. 32, no. 6, pp. 1065--1082, Jun. 2014.
\bibitem{6G} K. B. Letaief, W. Chen, Y. Shi, J. Zhang, and Y. A. Zhang, ``The roadmap to 6G: AI empowered wireless networks,'' {\it IEEE Commun. Magazine}, vol. 57, no. 8, pp. 84--90, Aug. 2019.
\bibitem{small_cells} V. Chandrasekhar, J. Andrews, and A. Gatherer, ``Femtocell networks: A survey,'' {\it IEEE Commun. Mag.}, vol. 46, no. 9, pp. 59--67, Sep. 2008.
\bibitem{D2D} D. Feng, L. Lu, Y. Yuan-Wu, G. Y. Li, S. Li, and G. Feng, ``Device-to-device communications in cellular networks,''  {\it IEEE Commun. Mag.}, vol. 52, no. 4, pp. 49-55, Apr. 2014.
\bibitem{Co_beam_2010} H. Dahrouj and W. Yu, ``Coordinated beamforming for the multicell multi-antenna wireless system,'' {\it IEEE Trans. Wireless Commun.}, vol. 9, no. 5, pp. 1748--1759, May 2010.
\bibitem{Co_beam_2011} Y.-F. Liu, Y.-H. Dai, and Z.-Q. Luo, ``Coordinated beamforming for MISO interference channel: Complexity analysis and efficient algorithms,'' {\it IEEE Trans. Signal Process.}, vol. 59, no. 3, pp. 1142--1157, Mar. 2011.
\bibitem{ICIC} R. Irmer, H. Droste, P. Marsch, M. Grieger, G. Fettweis, S. Brueck, H.P. Mayer, L. Thiele, and V. Jungnickel, ``Coordinated multipoint: Concepts, performance, and field trial results,'' {\it IEEE Commun. Mag.}, vol. 49, no. 2, pp. 102--111, Feb. 2011.
\bibitem{Net_MIMO3} Y.-F. Liu, Y.-H. Dai, and Z.-Q. Luo, ``Max-min fairness linear transceiver design for a multi-user MIMO interference channel,'' {\it IEEE Trans. Signal Process.}, vol. 61, no. 9, pp. 2413--2423, May 2013.
\bibitem{Net_MIMO1} J. Zhang, R. Chen, J.  Andrews, A. Ghosh, and R. Heath, ``Networked MIMO with clustered linear precoding,'' {\it IEEE Trans. Wireless Commun.}, vol. 8, no. 4, pp. 1910--1921, Apr. 2009.
\bibitem{Net_MIMO2} R. Zhang, ``Cooperative multi--cell block diagonalization with per-base-station power constraints,'' {\it IEEE J. Sel. Areas in Commun.}, vol. 28, no. 9, pp. 1435--1445, Dec. 2010.
\bibitem{Net_MIMO4} M. Hong, Q. Li, and Y. Liu, ``Decomposition by successive convex approximation: A unifying approach for linear transceiver design in heterogeneous networks," {\it IEEE Trans. Wireless Commun.}, vol. 15, no. 2, pp. 1377--1392, Feb. 2016.
\bibitem{IRS_wu} Q. Wu and R. Zhang, ``Towards smart and reconfigurable environment: Intelligent reflecting surface aided wireless network," {\it IEEE Commun. Mag.}, vol. 58, no. 1, pp. 106--112, Jan. 2020.
\bibitem{IRS_survey} E. Basar, M. Di Renzo, J. De Rosny, M. Debbah, M. Alouini, and R. Zhang, ``Wireless communications through reconfigurable intelligent surfaces," {\it IEEE Access}, vol. 7, pp. 116753--116773, Aug. 2019.
\bibitem{IRS_single1} Q. Wu and R. Zhang, ``Intelligent reflecting surface enhanced wireless network: Joint active and passive beamforming design,'' in {\it Proc. IEEE Globlecom}, 2018, pp. 1--6, Dec. 2018.
\bibitem{IRS_single2} X. Yu, D. Xu, and R. Schober, ``MISO wireless communication systems via intelligent reflecting surfaces : (Invited paper)," {\it 2019 IEEE/CIC ICCC}, Changchun, China, pp. 735--740, Aug. 2019.
\bibitem{IRS_multiuser} Q. Wu and R. Zhang, ``Intelligent reflecting surface enhanced wireless network via joint active and passive beamforming," {\it IEEE Trans. Wireless Commun.}, vol. 18, no. 11, pp. 5394--5409, Nov. 2019.
\bibitem{IRS_OFDM_protocol} Y. Yang, B. Zheng, S. Zhang, and R. Zhang, ``Intelligent reflecting surface meets OFDM: Protocol design and rate maximization.'' [Online] Available: {\url{https://arxiv.org/abs/1906.09956}}.
\bibitem{IRS_OFDM_est} B. Zheng and R. Zhang, ``Intelligent reflecting surface--enhanced OFDM: Channel estimation and reflection optimization,'' {\it IEEE Wireless Commun. Letters}., pp. 1--1, Dec. 2019.
\bibitem{IRS_OFDM} Y. Yang, S. Zhang, and R. Zhang, ``IRS--enhanced OFDM: Power allocation and passive array optimization,'' in {\it Proc. IEEE Globecom}, Dec. 2019. [Online] Available: {\url{https://arxiv.org/abs/1905.00604}}.
\bibitem{IRS_NOMA1} M. Fu, Y. Zhou, and Y. Shi, ``Intelligent reflecting surface for downlink non-orthogonal multiple access networks,'' in {\it Proc. IEEE Globecom}, Dec. 2019. [Online] Available: {\url{https://arxiv.org/abs/1906.09434}}.
\bibitem{IRS_NOMA2} G. Yang, X. Xu, and Y. C. Liang, ``Intelligent reflecting surface assisted non--orthogonal multiple access.'' [Online] Available: {\url{https://arxiv.org/abs/1907.03133}}.
\bibitem{SWIPT} Y. Tang, G. Ma, H. Xie, J. Xu, and X. Han, ``Joint transmit and reflective beamforming design for IRS-assisted multiuser MISO SWIPT systems,'' to appear in {\it Proc. IEEE ICC 2020.} [Online] Available: \url{https://arxiv.org/pdf/1910.07156}.
\bibitem{IRS_multicell}C. Pan, H. Ren, K. Wang, W. Xu, M. Elkashlan, A. Nallanathan, and L. Hanzo, ``Intelligent reflecting surface for multicell MIMO communications.'' [Online] Available: {\url{https://arxiv.org/abs/1907.10864}}.

\bibitem{SOCP} A. Wiesel, Y. C. Eldar, and S. Shamai, ``Linear precoding via conic optimization for fixed MIMO receivers,'' {\it IEEE Trans. Signal Process.}, vol. 54, no. 1, pp. 161--176, Jan. 2006.
\bibitem{Random} Z. Luo, W. Ma, A. M. So, Y. Ye, and S. Zhang, ``Semidefinite relaxation of quadratic optimization problems,'' {\it IEEE Signal Process. Mag.}, vol. 27, no. 3, pp. 20--34, May 2010.
\bibitem{CVX} M. Grant and S. Boyd, ``CVX: MATLAB software for disciplined convex programming,'' 2016. [Online] Available: {\url{http://cvxr.com/cvx}}.

\end{thebibliography}
\end{document}